\documentclass[10pt,aps,prl,twocolumn,preprint]{revtex4}   % style for Physical Review B and AJP are similar

\usepackage{amsmath}    % need for subequations
\usepackage{graphicx}   % for figures

\begin{document}

\title{The Cabibbo-Kobayashi-Maskawa density matrices}
\author{Miguel C. N. Fiolhais}
\email{miguel.fiolhais@cern.ch}   %optional
\affiliation{LIP, Department of Physics, University of Coimbra, 3004-516 Coimbra, Portugal}

\date{\today}

\begin{abstract}
The flavor changing charged currents of the weak sector of the Standard Model are studied in the framework of a quantum statistical approach. The quantum superposition of same-type quarks, generated by the Cabibbo-Kobayashi-Maskawa matrix, is used to define three density matrices, one for each quark family. The properties of such density matrices are analyzed, in particular, the associated von Neumann entropy. It is proven that, due to the unitarity of the Cabibbo-Kobayashi-Maskawa matrix, the quantum mixtures of quarks resulting from the weak interaction do not increase entropy and, therefore, the violation of CP and T symmetries cannot be related to the second law of thermodynamics.

The following article is published in Europhysics Letters: http://iopscience.iop.org/0295-5075/98/5/51001.
\end{abstract}

\maketitle

\section{Introduction}

It is now almost fifty years since the publication of Nicola Cabibbo's famous letter ``Unitary Symmetry and Leptonic Decays'' \cite{cabibbo}, which provided the basic foundation for quark mixing in the weak interaction sector of the Standard Model. Ten years later, Makoto Kobayashi and Toshihide Maskawa extended the work of Cabibbo to three generations of quarks and established the unitary Cabibbo-Kobayashi-Maskawa (CKM) matrix \cite{CKM}. The work of Kobayashi and Maskawa became notorious for explaining the observed CP violation \cite{christenson} in the context of a renormalizable theory of the weak interactions, and for predicting the existence of the bottom and top quarks \cite{herb,d0,cdf}, the third generation of quarks. As a result and according to the CPT theorem \cite{schwinger,schwinger2,schwinger3,schwinger4,luders,pauli,jost}, the weak sector of the Standard Model was settled in its present form, violating both CP and T symmetries by the same amount, in order to preserve CPT invariance.

The quarks charged current sector in the Standard Model Lagrangian, \emph{i.e.} the V-A coupling to the $W$ boson, can be written in terms of mass eigenstates as

\begin{eqnarray}
\mathcal{L}_{CC} = -\frac{g}{2\sqrt{2}} W_\mu^\dagger \sum_{i,j} \bar{u}_i \gamma^\mu (1-\gamma_5) \textbf{\textrm{V}}_{ij}d_j + h.c.,
\end{eqnarray}
where $i$ and $j$ are the quark generation indices and $\textbf{\textrm{V}}$ is the CKM matrix. The down-type weak eigenstates, $| d'_i \rangle $, are related to the mass eigenstates, $| d_j \rangle $, through
\begin{eqnarray}
\label{linearcomb}
| d'_i \rangle = \sum_{j} \textbf{\textrm{V}}_{ij} | d_j \rangle,
\end{eqnarray}
so that each up-type quark couples with every down-type quark by means of the CKM matrix. The CKM matrix is a $3\times3$ unitary matrix with four independent parameters. The unitarity constraints are expressed by
\begin{eqnarray}
\label{uni1}
\sum_{i} |\textbf{\textrm{V}}_{ij}|^2 = \sum_{j}|\textbf{\textrm{V}}_{ij}|^2 = 1,
\end{eqnarray}
for each quark generation, and by the triangular relations
\begin{eqnarray}
\label{uni2}
\sum_{k} \textbf{\textrm{V}}_{ik} \textbf{\textrm{V}}_{jk}^* = 0,
\end{eqnarray}
for any $i\neq j$. The CKM matrix can be parameterized in many different ways \cite{wolfenstein,chau}. The standard parameterization of the CKM matrix, used in this letter, comprises three mixing angles $(\theta_{12},\theta_{23},\theta_{13})$ and a CP-violating phase $\delta_{13}$, where $\theta_{12}$ represents the Cabibbo angle.

The CKM matrix is nowadays one of the pillars of the Standard Model of particle physics and has been experimentally scrutinized, along the years, at the main high-energy facilities, such as the Large Electron-Positron Collider, the Tevatron, the Brookhaven National Laboratory, the SLAC National Accelerator Laboratory and, more recently, at the Large Hadron Collider, among others \cite{pdg}. In this letter, the mixing of quarks is analyzed in the scope of quantum statistical mechanics, following a similar approach first used by Miller \cite{miller} for $SU(3)_c$ quark states. The density matrices built from the superposition of quarks of different generations are used to calculate the von Neumann entropy and conclusions are drawn on the CP and T symmetry violations and on the absence of an arrow of time in particle physics.

\section{Time asymmetries}

The different time asymmetries can be classified in three categories \cite{quinn}: the universal t-asymmetry, related to the expansion of the universe; the macroscopic t-asymmetry, also known as the second law of thermodynamics, in which the arrow of time is associated with the increase of entropy; and the microscopic t-asymmetry, present in several particle physics theories such as the Standard Model. The different time asymmetries are not related to each other, so the microscopic t-asymmetry cannot provide a microscopic explanation of the second law of thermodynamics observed at a macroscopic level. Even in the weak interaction of the Standard model, where CP and T symmetries are violated, the implications are that the CP eigenstates of a particle, for example the neutral kaon, have different masses and different decay widths, a fact that, by itself, does not imply any arrow of time. In fact, any microscopic physics process, when observed backwards in time, is perfectly symmetric and does not violate any fundamental physical law, even if it is statistically unlikely. However, despite the unidirectionality of time is only determined by the increase of entropy, this question has never been addressed, in such terms, in particle physics.

\section{CKM density matrices}

The density matrix, independently introduced by John von Neumann \cite{vonneumann} and by Lev Landau \cite{landau}, in 1927, is a self-adjoint positive-semidefinite matrix of trace one, particularly useful to describe and perform calculations on mixed states, $\emph{i.e.}$ linear combinations of several quantum states. For a superposition of quantum states,
\begin{eqnarray}
| \Psi \rangle = \sum_{n}  C_n | \Psi_n \rangle,
\end{eqnarray}
the density matrix is defined as
\begin{eqnarray}
\rho = \sum_{n,m}  C_n C_m^* | \Psi_n \rangle \langle \Psi_m |.
\end{eqnarray}
Accordingly, three CKM density matrices can be constructed from the quark mixing presented in eq. (\ref{linearcomb}),
\begin{eqnarray}
\rho_i = \sum_{j,k} \textbf{\textrm{V}}_{ij} \textbf{\textrm{V}}_{ik}^* | d_j \rangle \langle d_k |,
\end{eqnarray}
one for each generation of quarks. The hermiticity and trace one are clearly visible, for example, in the density matrix for the first generation which reads

{\small
\begin{equation}
\rho_1 =
\left( {\begin{array}{ccc}
 c_{12}^2c_{13}^2 & c_{12}s_{12}c_{13}^2 & c_{12}c_{13}s_{13}e^{i\delta_{13}} \\
 c_{12}s_{12}c_{13}^2 & s_{12}^2c_{13}^2 & c_{13}s_{12}s_{13}e^{i\delta_{13}} \\
 c_{12}c_{13}s_{13}e^{-i\delta_{13}} & c_{13}s_{12}s_{13}e^{-i\delta_{13}} &s_{13}^2 \\
 \end{array} } \right),
\end{equation}}where $c_{ij} = \cos \theta_{ij}$ and $s_{ij} = \sin \theta_{ij}$. The three density matrices are singular (non-invertible) and share one additional and peculiar property: each one of the three matrices has only one non-zero eigenvalue, which is equal to one. This queer feature appears as a result of the unitarity triangles of the CKM matrix given by eq. (\ref{uni2}), and has a direct impact on the von Neumann entropy.

The von Neumann entropy \cite{vonneumann2,landau2,landau3} is the quantum generalization of the classical entropy and, for a given density matrix, $\rho$, its definition is
\begin{eqnarray}
S(\rho) = -\mbox{Tr} \left ( \rho \log \rho \right ),
\end{eqnarray}
or, in terms of the $\rho$ matrix eigenvalues,
\begin{eqnarray}
S(\rho) = - \sum_{j} \lambda_j \log \lambda_j.
\end{eqnarray}
As the only non-zero eigenvalue of each density matrix is equal to one and since
\begin{equation}
\lim_{x \to 0} x \log x = 0,
\end{equation}
the von Neumann entropy is zero for all the three density matrices. As a result, the quark mixing does not increase the entropy, a direct consequence of the unitarity of the CKM matrix.

\section{Discussion}

The quark mixing density matrices were defined and analyzed in this letter. Charged currents in the weak sector of the Standard Model lead to quark mixing, but one concludes that such processes do not increase the von Neumann entropy. Despite this result might seem unnatural, since quantum mixed states with zero entropy are not usually found, it makes perfect sense from the physical point of view. Even though the quark mixing violates CP symmetry and, therefore, the T symmetry to preserve the CPT invariance, neither CP nor T violations represent any arrow of time, so entropy should not be increased. In other words, T symmetry violation in the dynamical laws of particle physics is not related to the second law of thermodynamics. Since the result obtained here is primarily due to the unitarity of the CKM matrix, it constraints possible new physics in quark flavor mixing beyond the Standard Model of particle physics. The non-unitarity of the CKM matrix would imply an increase of entropy, and therefore, an arrow of time.


\begin{thebibliography}{5}

\bibitem{cabibbo} N. Cabibbo, Phys. Rev. Lett.  \textbf{10}, 531 (1963).
\bibitem{CKM} M. Kobayashi and T. Maskawa, Prog. Theor. Phys. \textbf{49}, 652 (1973).
\bibitem{christenson} J. H. Christenson et al., Phys. Rev. Lett. \textbf{13}, 138 (1964).
\bibitem{herb} S. W. Herb et al.,  Phys. Rev. Lett. \textbf{39}, 252 (1977).
\bibitem{d0} S. Abachi et al. (D$\emptyset$ Collaboration), Phys. Rev. Lett. \textbf{74}, 2422 (1995).
\bibitem{cdf} F. Abe et al. (CDF Collaboration), Phys. Rev. Lett. \textbf{74}, 2626 (1995).
\bibitem{schwinger} J. Schwinger, Phys. Rev. \textbf{82}, 914 (1951).
\bibitem{schwinger2} J. Schwinger, Phys. Rev. \textbf{91}, 713 (1953).
\bibitem{schwinger3} J. Schwinger, Phys. Rev. \textbf{91}, 728 (1953).
\bibitem{schwinger4} J. Schwinger, Phys. Rev. \textbf{94}, 1362 (1954).
\bibitem{luders} G. L\"{u}ders, Dansk. Mat. Fys. Medd.  \textbf{28}, 5 (1954).
\bibitem{pauli}W. Pauli, \textsl{Exclusion principle, Lorentz group and reflexion of space-time and charge} (Pergamon Press, London, 1955).
\bibitem{jost} R. Jost, Helv. Phys. Acta \textbf{30}, 409 (1957).
\bibitem{wolfenstein} L. Wolfenstein, Phys. Rev. Lett. \textbf{51}, 1945 (1983).
\bibitem{chau} L.L. Chau and W.-Y. Keung, Phys. Rev. Lett. \textbf{53}, 1802 (1984).
\bibitem{pdg} K. Nakamura et al. (Particle Data Group), J. Phys. G \textbf{37}, 075021 (2010).
\bibitem{miller} D. E. Miller, Eur. Phys. J. C \textbf{34}, 435 (2004).
\bibitem{quinn} H. R. Quinn, J. Phys.: Conf. Ser. \textbf{171}, 012001 (2009).
\bibitem{eddington} A.S. Eddington, \textsl{The Nature of the Physical World} (The Macmillan Company, London, 1928).
\bibitem{vonneumann} J. von Neumann, G\"ottinger Nachrichten \textbf{1}, 245 (1927).
\bibitem{landau} L.D. Landau, Zeitschrift f\"ur Physik \textbf{45}, 430 (1927).
\bibitem{vonneumann2} J. von Neumann, \textsl{Mathematische Grundlagen der Quantenmechanik} (Springer Verlag, Berlin, 1932).
\bibitem{landau2} L.D. Landau and E.M. Lifshitz, \textsl{Course of theoretical physics, Vol. 3, Quantum mechanics} (Pergamon Press, Oxford, 1987).
\bibitem{landau3} L.D. Landau and E.M. Lifshitz, \textsl{Course of theoretical physics, Vol. 5, Statistical physics} (Pergamon Press, Oxford, 1985).

\end{thebibliography}
\end{document}